\documentclass[prd,letterpaper,twocolumn,tightenlines,superscriptaddress,
showpacs,psfig,showpacs,showkeys]{revtex4}

\usepackage[english]{babel} % Specify a different language here - english by 
\usepackage{lipsum} % Required to insert dummy text. To be removed otherwise

\usepackage{amsmath}
\usepackage{amsfonts}
\usepackage{amssymb}
\usepackage{amsthm}
\usepackage{fontenc}
\usepackage{graphicx}
\usepackage{tikz}
\usepackage{booktabs}
\usepackage{epsfig}
\usepackage{color}
\usepackage{ulem}
\usepackage{multirow} %this is for multiple rows you don't need if you wont use 
\usepackage{comment}
\newcommand\ddfrac[2]{\frac{\displaystyle #1}{\displaystyle #2} }

\usepackage{hyperref} % Required for hyperlinks
\begin{document}

%\pacs{}
%\keywords{}

\title{Hadronic structure from double parton scattering } % 
%Article title

\author{Matteo Rinaldi}\email{mrinaldi@ific.uv.es}
\affiliation{Departamento de F\'isica Teo\'orica-IFIC, Universidad 
de Valencia- CSIC, 46100 Burjassot (Valencia), Spain}
\author{Federico 
Alberto Ceccopieri}
\affiliation{IFPA, Universit\'e de Li\`ege,  B4000, Li\`ege, Belgium}

%\Authors{Matteo Rinaldi\textsuperscript{1}*,Federico Alberto Ceccopieri\textsuperscript{2}} 
%\affiliation{\textsuperscript{1}\textit{Departamento de F\'isica 
%Teo\'orica-IFIC, Universidad de Valencia- CSIC, 46100 Burjassot (Valencia), Spain}} 
%\affiliation{\textsuperscript{2}\textit{IFPA, Universit\'e de L\`iege, B4000 L\`iege, Belgium}} % Author affiliation
%\affiliation{*\textbf{Corresponding author}: mrinaldi@ific.uv.es} 

%\Keywords{} 
%\newcommand{\keywordname}{Keywords} % Defines the keywords heading name

%-------------------------------------------------------------------------------

%	ABSTRACT
%-------------------------------------------------------------------------------
\begin{abstract}
\noindent
%  \Abstract{
In the present paper we consider the so-called effective cross section, a 
quantity which encodes the  experimental knowledge on double 
parton scattering in hadronic collisions that has been accomulated so far. 
We show that the effective cross section, under some assumptions close to those 
adopted in its experimental extractions, can be used  to obtain a range of 
mean transverse distance between an interacting parton pair in double parton 
scattering. Therefore we have proved that the effective cross section offers a 
way to access  information on the hadronic structure.
%}
\end{abstract}

%-------------------------------------------------------------------------------
%\keywords{Double parton scattering, hadron collider}
\maketitle

%\begin{document}

%\flushbottom % Makes all text pages the same height

%\maketitle % Print the title and abstract box

%\tableofcontents % Print the contents section

%\thispagestyle{empty} % Removes page numbering from the first page

 \section{Introduction}
A proper description of the event structure in hadronic collisions requires 
the inclusion of the so-called multiple parton interactions (MPI) {which affect 
both the   multiplicity and topology of the hadronic final 
state\cite{paver,Sjostrand:1986ep}.} 
The large hadron collider operation 
renewed the interest in MPI given the continuous demand for an increasingly 
detailed description of the hadronic final state, which is crucial in many 
new physics searches.
In this rapidly evolving context, these types of studies have received attention
for their own sake: they might be sensitive to partonic correlations in the 
colliding hadrons. 
The simplest MPI process is double parton scattering (DPS). 
In such a case, a large momentum transfer is involved  {in both scatterings} 
and perturbative techniques can be applied to calculate the corresponding cross 
section. 
The latter depends  on a  two-body 
nonperturbative quantity,  the so-called double parton distribution 
functions (dPDFs).  
These distributions are interpreted as  number densities of  parton pairs  
with a given  transverse distance, $b_\perp$, in coordinate space and carrying  
longitudinal momentum fractions 
($ x_1,x_2$) of the parent proton \cite{ww_3_1}.
%\begin{align}
%\label{vera} 
% &  \hskip -0.9cm F_{ij}(x_1,x_2,{\vec z}_\perp)  = 
%(-8 \pi P^{+}) 
%\int \left [ \underset{l}{\overset{3}\prod}  \dfrac{
%d z_l^-}{4 \pi} \right]e^{i \frac{P^+}{2}(x_1 z_1^-+x_2 %z_2^--x_1 z_3^- )}
%\\
%\nonumber
%& \times 
%\langle  \vec P   \big|
%\hat{\mathcal{O}}_i^1 \left( z_1^-\dfrac{\bar n}{2},z_3^- %\dfrac{\bar n}{2}+
%\vec z_\perp \right)
%\hat{\mathcal O}_j^2 \left( z_2^-\dfrac{\bar n}{2}+
%\vec z_\perp,0 \right)
%\big| P  \rangle  ~.
%\end{align}
%The light-like four vector,
%$\bar n  = (1,0,0,-1)$, and 
%the nucleon state,
%$\big|  P  \rangle$ of momentum $P^\mu$,
%have been introduced. Furthermore~\\ $\hat{\mathcal O}_i^k(z,z') = 
%\bar q_i(z) \Gamma/2 q_i(z')$, where $\Gamma=\{\gamma^+ ,\gamma^+\gamma_5, 
%i\sigma^{j+}\gamma_5  \}$ and $i,j$ are flavor indexes. 
%Here and in the following,
%the ``$\pm$''  component of a vector defines the following combination $b^\pm = 
%b_0 \pm b_z$ and 
%$x_i = {k_i^+}/{P^+}$. 
% * <federico.alberto.ceccopieri@cern.ch> 2017-11-26T10:18:32.655Z:
% 
% Ho tolto il correlatore ed la sua discussione perchè non lo usiamo mai e non è bello vedere una formula nell'intro
% 
% ^.
Double PDFs are not calculable from first principles, a feature shared with
ordinary PDFs and other nonperturbative quantities in QCD. However, due to 
their  
dependence upon the partonic interdistance \cite{Calucci:1999yz},
they contain information on the hadronic structure    
complementary to those obtained from one-body distributions such as
generalized parton distribution functions (GPDs) and transverse momentum 
dependent PDFs. Unfortunately, since the DPS cross section depends on an 
integral
 over $b_\perp$, there are no 
experimental observables which may give direct access to such a 
dependence \cite{paver}. 

In this scenario, calculations of dPDFs via hadronic models have been used to 
obtain basic information and to gauge the impact of longitudinal and transverse 
correlations \cite{Mel_19,noiold,noi1,noij2,noir}. 
Despite this wealth of information possibly encoded in dPDFs, the  
experimental knowledge on DPS cross section has been accumulated, up to now, 
into the 
so-called effective cross section, $\sigma_{eff}$.  
The latter is defined through the ratio of the product of  two single 
parton scattering 
cross sections to the DPS cross section with the  same final states. 
The effective cross section 
has been extracted, although in a model dependent way, in several experiments; 
see recent results in Refs. \cite{data6,data8,data9,data10,data11,data12}. 
%\sout{ The 
%conclusion, within the accuracy of the data and under the assumption of independent hard scatterings, is that $\sigma_{eff}$  is compatible with a constant and 
%it does  depend neither on the center-of-mass energy of the collision nor on the type and kinematics of the produced final state.} 
The purpose of the present paper is to demonstrate that, by exploiting the
maximum information encoded in $\sigma_{eff}$  
and by using
almost the same assumptions used in its experimental extraction, a range of
mean distances, characterizing the interacting parton pair, can be
derived. Thanks to this result, one can access  information on the hadronic
structure, encoded in the dPDFs $b_\perp$ dependence, in a quite  rather easy 
%\sout{and almost model independent} 
way without any detailed  knowledge on such a dependence in transverse space.
%In this sense, \textcolor{blue}{results here discussed can be considered as model independent.
Therefore,  the present analysis represents an attempt to generalize historical 
studies on the proton radius in exclusive processes to the relative partonic
distance between two interacting partons in DPS processes in hadronic 
collisions.

This paper is organized as follows. In Sec. \ref{tomo}, we show how, in 
principle,
novel information on the proton structure  
can be achieved by means of dPDFs and  a new `` form factor'' is introduced.
In Sec. \ref{sigma_eff},
we elaborate a general relation between $\sigma_{eff}$ and the mean distance 
of two 
interacting partons.  In Secs. \ref{minimum} and \ref{maximum}, we derive a 
couple of  inequalities suitable  to extract information on the mean partonic 
distance from experimental values of
$\sigma_{eff}$.
In Sec. \ref{examples}, we discuss numerical results. We collect our conclusions
in Sec. \ref{conclusions}.

\section{Hadron structure via DPS }
\label{tomo}
Similarly to the case of GPDs, whose first moment is related to
standard Dirac and Pauli form 
factors, we may introduce the 
first moment of dPDFs with respect to $x_1$ and $x_2$
\begin{align}
\label{uno}
 f_{ij}(k_\perp)= \frac{1}{N_{ij}} \int dx_1~dx_2~  F_{ij}(x_1,x_2,k_\perp)~,
\end{align}
%\begin{align}
% f_{ij}(k_\perp)= \frac{1}{N_{ij}} \int dx_1~dx_2~  F_{ij}(x_1,x_2,k_\perp)~,
%\end{align}
where $i$ and $j$ are
%, \textsl{e.g.}, valence quarks
parton indices and we address $f_{ij}(k_\perp)$ as the ``effective 
form factor'' \cite{noiplb1}.
The functions $F_{ij}(x_1,x_2,k_\perp)$ are the Fourier transform of dPDFs 
$\tilde F_{ij}(x_1,x_2,b_\perp)$. 
According to Ref. \cite{noi22}, 
the 
%$1/2$ 
$N_{ij}$ factors in Eq. (\ref{uno}) represent the dPDF normalizations  
evaluated at $k_\perp=0$, e.g. for valence quarks $N_{u_v d_v}=N_{u_v u_v}=2$.
%The indexes $i,j$ indicate the
%partonic flavor. 
%and $N_{ij}$ is the normalization of the dPDFS, i.e. $ \int dx_1dx_2 F_{ij}(x_1,x_2,k_\perp=0)= N_{ij}$ \cite{noi22}. 
At variance with the GPD case, here $k_\perp$ does not represent a 
momentum transfer between the proton initial 
and final state but a rather {transverse momentum} imbalance between 
two partons in the amplitude and its conjugate \cite{ww_25_1}.
{Therefore in momentum space $F_{ij}(x_1,x_2,k_\perp)$ does not admit a 
probabilistic interpretation, which {holds} instead in $b_\perp$ space.
The effective form factor can be defined in a more fundamental manner in terms 
of the proton  wave function. In fact, in the nonrelativistic limit, it 
is given by
\begin{equation}
 f_{ij}(k_\perp)=
 \int d\vec k_1 d\vec k_2~ \Psi^\dagger(\vec k_1 +\vec k_\perp, \vec 
k_2) \tau_i \tau_j
\Psi(\vec k_1, \vec k_2 + \vec k_\perp)~,
\label{efff}
\end{equation}
   with   $\Psi(\vec k_1,\vec k_2)$ being the canonical proton wave function in 
 the 
intrinsic frame 
depending on the parton momentum $\vec k_i$ and $\tau_i$ the usual flavor 
projector, see, e.g., Ref. \cite{noiold}. The effective form factor can be 
related to the two-body density of partons,  
$\tilde f_{ij}(b_\perp)$,  with $b_\perp$ being the relative distance 
between  two 
partons, defined  
by means 
of the Fourier transform of the proton wave function with respect to $\vec 
k_\perp$, i.e.,   
%$\rho(\vec r_1,\vec r_2)$, defined  
%by means 
%of the Fourier transform of the proton wave function w.r.t. $\vec k_i$, i.e.:

\begin{align}
f_{ij}(k_\perp) = \int d\vec b_\perp\ ~
e^{i \vec k_\perp \cdot \vec b_\perp}\tilde f_{ij}(b_\perp)~.
 \label{efff2}
\end{align}
 Equations (\ref{efff}) and (\ref{efff2}) are similar to 
those used to define the standard electromagnetic proton form factor  
in terms of the same hadron wave function $\Psi(\vec k_1,\vec k_2)$.
Analogously to this standard case, one can define the mean value of the 
distance  between two partons
in the transverse plane  
through the effective form factor,
\begin{eqnarray}
 \langle b^2 \rangle_{ij}  \simeq 
-4\dfrac{d
~f_{ij}(k_\perp)}{d~k_\perp^2}\Bigg |_{k_\perp =0}.  
\label{d1}
\end{eqnarray}
The knowledge of $f_{ij}(k_\perp)$ gives access to new information, 
generalizing the results   on the proton mean radius, obtained from 
electromagnetic proton form factors in elastic processes.
Despite the richness of information encoded in  %\cite{Calucci:1999yz},
the effective form factor, 
this quantity is actually poorly known from the theoretical and experimental
points of view. In fact,
in DPS processes, only information on the integral of dPDFs with respect to
$k_\perp$ is available \cite{paver}.
In order to overcome this problem, in the next sections 
we present a procedure which  relates the mean partonic distance between 
two partons directly to the experimentally extracted $\sigma_{eff}$.

\section{$\sigma_{eff}$ and partonic distances}
\label{sigma_eff}
The differential DPS cross section, assuming that the two hard scattering 
processes can be factorized \cite{ww_3_1,ww_25_1,Diehl:2015bca, add1,add2},  
involves dPDFs through an integral over $k_\perp$ and reads \cite{ww_25_1} 
\begin{multline}
 \label{sigma_DPS} 
 d\sigma_{DPS}^{A+B}= \frac{m}{2}  \int
  \frac{d^2k_{\perp}}{(2\pi)^2} \; d\hat{\sigma}_{ik}^A 
\; d\hat{\sigma}_{jl}^B \cdot \\
 \cdot F_{ij}(x_1,x_2,k_{\perp})
 F_{kl}(x_3,x_4,-k_{\perp})
 \,.
\end{multline}
It represents the 
Fourier-transformed version of the DPS cross section formula in $b_\perp$ space 
presented in  Ref. \cite{paver}.
In Eq. (\ref{sigma_DPS}) 
 $d\hat{\sigma}$ are the differential partonic cross sections for processes 
 A and B, respectively and the symmetry factor $m=1$ if $A=B$ and $m=2$ 
otherwise. %\textcolor{blue}{
%Since in present analysis the above cross section is integrated over $x_1$ and $x_2$ in order to have a significant statistic in order to observe DPS and considering the analytic structure of
% the above equation, a direct extraction of dPDFs from DPS is a theoretical challenge. However, as will be discussed later on some information on the proton structure can be still obtained from data.  }
%proton lost. However, as will be discussed later on, since a mean value of the DPS cross section has been obtained in different kinematic conditions, in the present section we show how some clue on the proton structure can be partially restored.
Given the limited knowledge regarding dPDFs, a fully factorized ansatz is 
frequently assumed:
\begin{align}
F_{ij}(x_1,x_2,k_\perp) \sim q_i(x_1)q_j(x_2)f(k_\perp)\,,
\label{ans}
\end{align}
where $q_i(x)$ are ordinary PDFs. 
%(\textcolor{blue}{Va bene, spostiamole  qui e magari mettiamone sono una o due, giusto per citare e per far sottilineare che questo ansatz è anche stato  utilizzato e leggermente motivato da "teorici" diciamo, tutto qui.}). 
Usually,  in such a simplified approach, the transverse form factor, 
$f(k_\perp)$,  depends neither on parton flavors nor on its fractional 
momenta \cite{Mekhfi:1983az}. It is worth mentioning that dPDF calculations 
within hadronic models show, in general, a breaking of the factorized ansatz, 
Eq. (\ref{ans}),  in a specific region of phase space, where sizable 
longitudinal 
and mixed longitudinal-transverse partonic correlations do appear 
\cite{Mel_19,noiold,noi1,noij2,noir}.
Nevertheless in this paper we still  use the approximation in Eq. (\ref{ans})
in 
order to make contact with experimental extractions of $\sigma_{eff}$. We 
remark, however, that in the present work, no
assumptions on the detailed functional form of $f(k_\perp)$ are used.   
In such a case 
 the DPS cross section simplifies to the form \cite{Calucci:1999yz}
\begin{equation}
 \label{sff}
 d\sigma_{DPS}^{A+B}  = \frac{m}{2}\dfrac{d\sigma_{SPS}^A d\sigma_{SPS}^B 
}{ \sigma_{eff}}~,
\end{equation}
  with $d\sigma_{SPS}^A{(B)}$ being the single parton scattering cross section 
 with
final state $A(B)$. 
% In particular, we discuss this topic considering results of analyses on 
%$\sigma_{eff}$. To this aim
%let as introduce $\sigma_{eff}$ 
%\cite{Mel_11,plb_5},  From experimental analyses 
%\cite{data6,data7,data8,data9,noiww,cao}, it has been found 
%that the mean value of the effective cross section is
%$\sigma_{eff}\sim 16$ mb.
%It is necessary to point out that in these experimental analyses, use 
%has been made of the following ansatz:
In this scenario, $\sigma_{eff}$ is simply given in $k_\perp$ space by
%\begin{align}
%\label{sap}
% \sigma_{eff}^{-1} = \int { d^2k_{\perp} \over (2\pi)^2} f(k_\perp)^2 = 
%\int_0^\infty 
% { dk_\perp \over 2\pi}~k_\perp f(k_\perp)^2~.
%\end{align}
\begin{align}
\label{sap}
 \sigma_{eff}^{-1} = \int { d^2k_{\perp} \over (2\pi)^2} f(k_\perp)^2~
{
=\int { d k_{\perp} \over 2\pi} k_{\perp} f(k_\perp)^2} ,
\end{align}
where the last expression follows from rotational invariance since we are 
interested in scattering processes whose final states are integrated over 
angles.
Equation (\ref{sff}) shows that $\sigma_{eff}$
enters the DPS cross section formula as an overall normalization factor. 
%\textcolor{blue}{The main purpouse of the present analysis is to use the 
%analythic structure of $\sigma_{eff}$, Eq. (\ref{sap}) togheter with 
%properties of the effective form factor to obtain new information on the 
%partonic proton structure from experimental analyses of $\sigma_{eff}$.}
%However, the approximation Eq. (\ref{ans}) requires two factorization ansatz, 
%i.e. the one in the $x_1-x_2$ dependence and the one in the 
%x_1,x_2)-k_\perp$ 
%dependence,  scenarios that can be violated due to different kind of partonic 
%correlations,see   
%Since, as already mentioned, experimental analyses provide useful 
%information on $\sigma_{eff}$, integral Eq. (\ref{sap}), in this paper we 
%study how this kind of outcomes can unveil some clues on the unknown 
%$f(k_\perp)$. 
Starting from Eq. (\ref{sap}), we show in this section how  such  an integral 
 can be 
related to the mean distance of the two partons involved in the scattering 
process.  
\textcolor{black}{For this purpose, we use two properties
granted from the general structure
of the hadronic wave function in Eq. (\ref{efff}), i.e., 
\begin{equation}
f(k_\perp=0)=1 \;\;\; \mbox{and} \;\;\; f(k_\perp \rightarrow\infty)=0.
\label{assumptions}
\end{equation}}
%To the  aim of the present paper, the integral in Eq.(\ref{sap}) has been  
%studied by considering only general model independent
%properties of $f(k_\perp)$, i.e. , addressed from Eq. (\ref{efff}). 
Thanks to the latter conditions, two identities are immediately obtained,
\begin{multline}
\label{id1g}
\hskip -0.7cm \int_0^\infty 
dk_\perp~k_\perp^m f(k_\perp)^2 =
\\
-2\int_0^\infty 
dk_\perp~   {k^{m+1}_\perp \over m+1 } f(k_\perp){d \over dk_\perp}f(k_\perp),
\end{multline}
with $m \geq 0$ and
%The latter is found by integrating by parts the left hand side 
%and properly taking into account the properties of $f(k_\perp)$ 
%already discussed.Moreover,
\begin{align}
\int_0^\infty 
dk_\perp~ f(k_\perp)^{s-1} {d \over dk_\perp}f(k_\perp)=-{f(0)^s \over s}=-{1 
\over s},
\label{id2}
\end{align}
which will be frequently used in the following.
Furthermore, with $\vec k_\perp$ being  defined on the transverse plane, in two 
dimensions,
$f(k_\perp)$ can be 
defined as 
%nn important ingredient 
%for the present purpose, is the expansion of the effective form factor 
%in terms 
%of powers of the main distance between partons, using Eq. (\ref{efff2}). 
%In fact, 
\begin{align}
{
 f(k_\perp) = \int d^2b_\perp~e^{i \vec k_\perp \cdot \vec b_\perp}
 \tilde f(b_\perp)= 2\pi
 \int db~\tilde f(b) J_{0}(k_\perp b),}
\end{align}
 with $\tilde f(b)$ being the  probability density of finding two partons 
 with a relative transverse distance $b=|\vec b_\perp|$,  $\tilde f(b)$ being a 
radial function of $b$. 
By expanding in series the Bessel 
function $J_0(k_\perp b )$ we find the following useful representation
\begin{align}
\label{34l}
 f(k_\perp) &= \sum_{n=0}^\infty {  k_\perp^{2n}  \langle b^{2n} \rangle 
 } \dfrac{ (-1)^n }{4^n   (n!)^2 }
=\sum_{n=0}^\infty {  k_\perp^{2n}  \langle b^{2n} \rangle 
 } P_n^{J_0}\,,
\end{align}
where the $P_n^{J_0}$ are the coefficients of the Bessel expansion and
$\langle b^{2n} \rangle$  are the $2n$ moments  of $\tilde f(b)$ 
and contain all dynamical unknown information on partonic proton structure.
At this point, we arrange Eq. (\ref{sap}) in a  form more suitable  
for our purposes.
We consider Eq. (\ref{id2}) for $s=3$ and, by using the expansion 
in Eq. (\ref{34l}) with the  $n=0$ and $n=1$ terms kept explicit, we get

\begin{align}
 & -{1 \over 3}=\int_0^\infty dk_\perp~f(k_\perp)^2 f'(k_\perp) =
\\
\nonumber
& \int_0^\infty dk_\perp~f(k_\perp) f'(k_\perp)\left[1- \dfrac{k_\perp^2 
\langle b^2 \rangle}{4} + \sum_{n=2}P_n^{J_0}
k_\perp^{2n}\langle b^{2n} \rangle \right]\,.
\end{align}
The terms in square brackets are then evaluated as follows. 
The first one is simplified by using 
Eq. (\ref{id2}) with $s=2$,  the second one by using Eq. (\ref{id1g})
with $m=1$, and the last term by using Eq.(\ref{id1g}) with $m=2n-1$. 
Collecting 
results and dividing by $\langle b^2\rangle/4$, we find:
%\begin{align}
% \hskip -0.8cm -{1\over6}+{\langle b^2\rangle  \over 4} \int_0^\infty 
%dk_\perp~k_\perp 
%f(k_\perp)^2\\
%\nonumber
%-\sum_{n=2} P_n^{j_0} n \langle b^{2n}\rangle  \int_0^\infty 
%dk_\perp~k_\perp^{2n-1} 
%f(k_\perp)^2 =0~.
%\end{align}

% Dividing for $\langle b^2\rangle/4$ one gets:
 
 \begin{align}
\label{f1}
 \hskip-0.6cm  &\int_0^\infty dk_\perp~k_\perp f(k_\perp)^2= 
 {2 \over 3 \langle b^2 \rangle  } +
\\
\nonumber
+
& 4 \sum_{n=2}  {{ \langle b^{2n} \rangle  }P_n^{J_0} ~n\over  \langle 
b^2
\rangle }
\int_0^\infty dk_\perp  k_\perp^{2n-1} f(k_\perp)^2. 
\end{align}
Although Eq. (\ref{f1}) shows a formal relation between $\sigma_{eff}$ and
$\langle b^2 \rangle$, the latter is obscured by the last term, which requires 
the explicit knowledge of $f(k_\perp)$. In the next two subsections we show how 
this problem can  actually be circumvented providing an easy-to-evaluate 
relation 
between $\sigma_{eff}$ and $\langle b^2 \rangle$.
We mention here for later convenience that 
by a repeated use of the  
Cauchy-Schwarz inequality and 
the property of the variance, $\langle b^2 \rangle  \geq  \langle b \rangle^2$,
it can be easily shown that 
 \begin{align}
 \langle b^{n } \rangle  \geq  \langle b \rangle^{n}\,,
\label{dx4b}
\end{align} 
which represents a generalization of the property of the variance for  
$n\geq 2$.

\subsection{A minimum for the allowed partonic distance} \label{minimum}
In this subsection we show how, 
given a known value for $\sigma_{eff}$,
a minimum value for the mean partonic distance can be derived by using 
Eq. (\ref{f1}).
%Motivated by the lack of clues on three dimensional structure of the proton in 
%terms of two partons, i.e. the knowledge of $\langle b^2\rangle$, in 
%this part we investigate how, from Eq. (\ref{f1}), a model independent  
%information on $\langle b^2\rangle$ can be obtained from the available
%experimental analyses of $\sigma_{eff}$.
For this purpose, generalizing Eq. (\ref{d1}), we introduce 
the function 
\begin{equation}
d_2(k_\perp) = -2 f'(k_\perp)/k_\perp\,.
\label{d2k}
\end{equation}
By using the expansion for $f(k_\perp)$ in Eq. (\ref{34l}), one finds
\begin{equation}
 d_2(k_\perp) = -4 \sum_{n=1} k_\perp^{2n-2} \langle b^{2n}\rangle
 P_n^{J_0}n = \langle b^{2}\rangle + \mathcal{O}(k_\perp^2), 
 \label{d2k_expansion}
\end{equation}
which immediately gives  $d_2(k_\perp=0)=\langle b^2\rangle$. 
At this point one may notice that the formal definition of $f(k_\perp)$, Eq. 
(\ref{efff}), is rather similar to the one of the electromagnetic proton 
form factor, 
except that in the present case $k_\perp$ is a transverse momentum imbalance
in a two-body distribution.
Since  electromagnetic proton form factors 
are, in general, decreasing functions of $k_\perp$, we may 
expect a similar behavior 
in $f(k_\perp)$. This observation implies that 
 $d_2(k_\perp) \geq 0$ via Eq. (\ref{d2k}).
Additionally, one may notice that 
\begin{align}
 \dfrac{d}{  k_\perp d ~k_\perp}d_2(k_\perp) \Bigg|_{k_\perp =0} =-8 
 P_2^{J_0}<0\,,
\end{align}
implying that 
$d_2(0)$ is a maximum for $d_2(k_\perp)$.
At this point, one may consider the identity in Eq. (\ref{id2}) with $s=3$,
\begin{align}
\label{eq17}
 \int_0^\infty dk_\perp~k_\perp f(k_\perp)^2 d_2(k_\perp) =  {2 / 3}.
\end{align}
Since $d_2(0)$ is a maximum for $d_2(k_\perp)$,
%we can 
%study the derivative of $d^2(k_\perp)$ to show that $\langle b^2\rangle 
%=d^2(0)$ 
%is the 
%maximum. 
%In fact, 
%\begin{align}
% \dfrac{d}{ d ~k_\perp}d^2(k_\perp) = -8 \sum_{n=2} k^{2n-3} \langle 
%b^{2n}\rangle P_n^{J_0}n(n -1)
%\end{align}
%now, for $k_\perp=0$ this function is zero. Moreover, in order to study the 
%sign of the function near the point $k_\perp \sim 0$, it is sufficient to 
%analyze the sign of the function
%Since   $P_2^{j_0} > 0$,  the above expression is negative, so that we deduce 
%that the derivative of $d^2(k_\perp)$ for $k_\perp \sim 0 $ is negative 
%\begin{align}
% \int_0^\infty dk_\perp~k_\perp f(k_\perp)^2 d^2(0) \geq  {2 \over 3}~, 
%\end{align}
% in other words:
we deduce from Eq. (\ref{eq17})
that 
\begin{align}
\int_0^\infty dk_\perp~k_\perp f(k_\perp)^2  \geq  {2 \over 3 \langle 
b^2\rangle},
\label{mag}
\end{align}
a result which can be rewritten in terms of the effective cross section as 
$\langle b^2\rangle \geq  \sigma_{eff}/(3\pi)$.
We remark that the same result can be obtained starting directly from
Eq. (\ref{f1}).
In fact, thanks to the variance property in Eq. (\ref{dx4b}) and the formal 
definition of $P_n^{J_0}$, one can analytically prove that the second term on
the right-hand side of  Eq. (\ref{f1}) is positive,  therefore leading to the 
same final result, Eq. (\ref{mag}).

\subsection{A  maximum for the allowed partonic distance }
\label{maximum}

In this subsection, we investigate whether $\sigma_{eff}$ determines a maximum 
value for the mean interpartonic distance. We note that the properties of 
$f(k_\perp)$ used up to now will not be sufficient for our purpose, and we will 
introduce additional reasonable assumptions which we will discuss during the 
proof.

%In particular, 
%let us remind the we can only use model independent properties of integrand of Eq. (\ref{sap}) 
% and identities like Eq. 
%(\ref{id1g},\ref{id2}). 

From the definition of $\sigma_{eff}$ in Eq. (8), we note that the integral is 
positive definite; thus, we can introduce an integer $\tilde N$ such that
\begin{align}
{ 2\pi \over \sigma_{eff} }= \int_0^\infty dk_\perp~k_\perp f(k_\perp)^2 = {1 
\over \tilde N \langle b^2 
\rangle }.
\label{sigma_eff_max32}
\end{align}
%By comparing with  Eq. (\ref{mag}) one gets the trivial solution $\tilde N <3/2$. Therefore , from the above identity,  
Therefore, for any  $N\leq \tilde N$,
\begin{align}
 \int_0^\infty dk_\perp~k_\perp f(k_\perp)^2 N \langle b^2\rangle  \leq 1.
 \label{s1}
\end{align}
Trivially, $N=0$ is a solution of this equation, which is of no interest.
However, given the result in Eq. (\ref{mag}), our problem reduces to the search 
of a nonzero value of $N$ in the range $0<N<3/2$. For this purpose, we  
subtract from Eq. (\ref{s1}) the  identity in Eq. 
(\ref{id2}) with  
$s=2$, obtaining
%\sout{ and $N=\tilde{N}=3/2$. 
%The latter value is obtained
%subtracting the  identity in Eq. (\ref{id2}) with  
%$s=3$ from  Eq. (\ref{s1}) and recalling that $ \langle b^2 \rangle  = d^2(k_\perp=0)$ is a maximum
%for $d^2$.} 
%\begin{align}
% \int_0^\infty dk_\perp~k_\perp f(k_\perp) d^2(k_\perp) = 1~.
% \label{s2}
%\end{align}
 %From Eqs. (\ref{s1},\ref{s2}) in principle we would need to find a range of 
%$N$ such that:
\begin{align}
  \int_0^\infty dk_\perp~k_\perp f(k_\perp)\Big[N \langle b^2 \rangle 
f(k_\perp)-d_2(k_\perp) \Big] 
\leq 0.
\label{s3}
\end{align}
%If such a range can be identified, than $1/\tilde N \leq 1/N\leq 1/ N_m$ and, in terms of effective cross 
%sections, we get the desired result:
%\begin{align}
%  \int_0^\infty dk_\perp~k_\perp f(k_\perp)^2 = {1 \over \tilde N} {1 \over 
%\langle b^2 \rangle} \leq  {1 \over N_m} {1 \over \langle 
%b^2 \rangle}~.
%\end{align}
%\begin{align}
% \langle b^2 \rangle  \leq {1 \over N_m} { \sigma_{eff} \over 2 \pi}~.
% \label{hope}
%\end{align}
%On the other hand by subtracting the  identity in Eq. (\ref{id2}) with  
%$s=3$, from Eq. (\ref{s1}), we obtain:
%\begin{align}
% \int_0^\infty dk_\perp~k_\perp f(k_\perp) d^2(k_\perp) = 1~.
% \label{s2}
%\end{align}
 %From Eqs. (\ref{s1},\ref{s2}) in principle we would need to find a range of 
%$N$ such that:
%\begin{align}
%  \int_0^\infty dk_\perp~k_\perp f(k_\perp)^2\Big[N \langle b^2 \rangle 
%-3/2 d^2(k_\perp) \Big] 
%\leq 0~.
%\label{s3bis}
%\end{align}
%\textcolor{blue}{The advantage of the above equation is in the chance to study the integrand in order to find a suitable solution. }
Finding  a solution to Eqs. (\ref{s1}) and (\ref{s3}) is not possible withouta
detailed knowledge of the functional form of $f(k_\perp)$. Nevertheless, we can 
study the sign of the term in square brackets in Eq. (\ref{s3}),  i.e., 
\begin{align}
 N\langle b^2 \rangle f(k_\perp) \leq d_2(k_\perp).
 \label{s33}
\end{align}
This inequality represents a sufficient condition for the validity of 
Eq. (\ref{s3}). The condition in not necessary because there might exist regions 
in $k_\perp$ and values of $N$ for which such a term is positive but the 
integral in Eq. (\ref{s3}) is negative.
To further proceed, let us rewrite Eq. (\ref{s33}) by using the series 
expansion of $f(k_\perp)$ and $d_2(k_\perp)$, obtaining
\begin{align}
 N \langle b^2 \rangle \sum_{n=0} P_n^{j_0}k_{\perp}^{2n}\langle b^{2n} 
 \rangle  \leq 
\sum_{n=0} { P_n^{j_0} \over n+1} k_{\perp}^{2n} \langle b^{2n+2} \rangle.
\end{align}
By equating terms of equal powers in $k_\perp$, we get the following 
set of solutions:
\begin{equation}
\frac{1}{n+1}<N<\frac{1}{n}, \;\;\;\;\;\; n=\mbox{odd}\,. 
\label{formal}
\end{equation}
Such  solutions, however, do not take into account the detailed $k_\perp-$
dependence of $f(k_\perp)$. For example, if the integral in Eq. 
(\ref{sigma_eff_max32}) is dominated by the low $k_\perp$ region, the solution 
to Eqs. (\ref{s3}) and (\ref{s33}) is found in the first interval, namely $1/2< 
N<1$.
Since this case corresponds to an effective form factor falling sufficiently
fast at large $k_\perp$, we take this condition as a working hypothesis and 
provide  supporting arguments in the following.

%Therefore, in order to provide a sensible value 
%for $N$ which ensures a realistic upper limit to $\langle b^2 \rangle$, we have to invoke additional assumptions on $f(k_\perp)$. 
%We therefore take  $N=1/2$ as a working hypothesis and provide physical arguments 
%to support it. 
%Moreover, by considering the asymptotic behavior , one can realize that such a procedure works fine fol all standard proton form factors
In the first place we wish to quantify 
the limiting asymptotics of $f(k_\perp)$ at large $k_\perp$,  
which satisfies the proposed solution.
For this purpose, we consider a dipole test function of the type
\begin{equation}
f(k_{\perp})=\Big(1+\frac{k_{\perp}^2}{m^2}\Big)^{-r}
\label{dipole}
\end{equation}
in which $m$ is a mass  parameter and the large $k_{\perp}$ behavior is 
controlled by the tunable parameter $r$.  
By direct evaluation, we find that our proposed solution is valid if $r>1$ in
Eq. (\ref{dipole}).
The same result holds  for  functions that fall even faster at large 
$k_{\perp}$ like Gaussians and exponentials. 

Secondly, additional support for the proposed solution is provided  by the 
following observation
\cite{strikman3} : 
$f(k_\perp)$ represents a two-body form factor,  $k_\perp$ being a transverse 
momentum imbalance between the parton pair. As such, its asymptotic behavior at 
large $k_{\perp}$ should 
fall  more rapidly than the one in one-body form factors. 
If one uses for  $f(k_\perp)$ 
the results obtained in Refs. \cite{em1,em2,em3}, one finds that the proposed 
solution is verified since these functions all have  dipole forms  with $r=2$. 
The same conclusion is reached if Dirac and Pauli form factors  are used, 
whose behaviors at large momentum transfer $Q$ are given by $1/Q^4$ and 
$1/Q^6$, 
respectively \cite{brod}. }
Finally, we remark that the proposed solution is found to be valid for model 
calculations of $f(k_\perp)$, in 
particular the one evaluated within the Light-Front approach in Ref. \cite{noi1}
and for 
the two-gluon form factor 
discussed in Ref. \cite{strikman3}. 
To conclude, we have found that Eq.  (\ref{s3}) is verified for $N=1/2$ under 
the additional condition that $f(k_\perp)$ falls off as $k_\perp^{-2}$ or 
faster.
As a consequence of our derivation
we can state that
\begin{align}
\int_0^\infty dk_\perp~k_\perp f(k_\perp)^2 \leq {2\over \langle b^2 \rangle }.
 \label{hoe2}
\end{align}
Combining this result with Eq. (\ref{mag}) leads to 
\begin{align}
  \dfrac{\sigma_{eff} }{3 \pi}  \leq \langle b^2 \rangle \leq   
\dfrac{\sigma_{eff} }{  \pi},
\label{main}
\end{align}
which limits the range of the interpartonic distance and is the main result of
 the paper.
\textcolor{black}{We wish 
to close this section highlighting the degree of model dependence of this 
result. 
The latter indeed does depend on the approximations made in 
Eqs.(\ref{sigma_DPS})
and (\ref{ans}), in particular on the full factorization  of $f(k_\perp)$ in 
the 
dPDF expression together with its flavor and energy dependence.
Therefore it contains the same model dependence assumed in  the $\sigma_{eff}$ 
extraction. 
However our result does depend weakly on the details of $f(k_\perp)$, 
since just the general conditions in Eq. (\ref{assumptions})
and its limiting asymptotics  at large $k_\perp$ 
are assumed, leaving the detailed shape $f(k_\perp)$ largely unconstrained.}

\begin{figure}
\begin{center}
% \special{psfile=dist_exp.eps hoffset=10 voffset=-150 hscale= 84 vscale=80}
%\includegraphics[scale=0.8]{dist_exp.eps}
\includegraphics[scale=0.65]{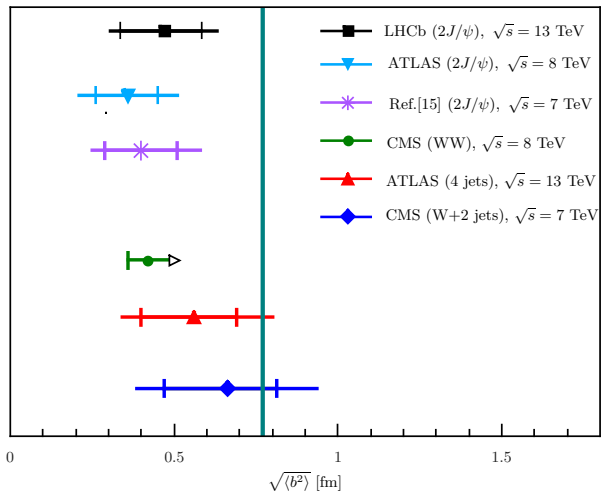}
%\vskip 5.5cm
\vskip 0.5cm
\caption{The range of allowed mean partonic distance, Eq. (\ref{main}), 
calculated by using $\sigma_{eff}$  extracted values from different  
experimental analyses \cite{data8,data9,data6,data10,data11,data12}. Inner error 
bars represent the theoretical uncertainty associated with the range in Eq. 
(\ref{main}). 
The outer ones represent the propagation of experimental uncertainties, 
related to $\sigma_{eff}$ extraction, plus theoretical ones added in 
quadrature. The vertical line represents the proton radius.}

\label{distexp}
\end{center}
\end{figure}

\section{Numerical results}
\label{examples}
In this section, we discuss  a direct application of Eq. (\ref{main}).
%\sout{  a couple of examples of the possible applications of Eq. (\ref{main}). In the first one
%we consider the theoretical model for the effective form factor proposed in Ref. \cite{strikman} }
%\begin{align}
%f(k_\perp) = \left( 1+{k_\perp^2}/{m_g^2}   \right)^{-4}~,
%\label{ffs}
%\end{align} 
%\sout{
%a form which is inspired by the gluon GPDs at small $x$, being $m_g^2=1.1$ Ge\mbox{$V^2$}. 
%Therefore we can evaluate explicitly Eqs. (\ref{d1},\ref{sap}) and obtain the values  $\sigma_{eff} 
%=28\pi/m_g^2$  and $\langle b^2\rangle= 16/m_g^2$.
%The latter results indeed satisfy Eq. (\ref{main}). }
Since the latter is derived with a set of assumptions  close  to the ones used 
by experimental collaborations to extract $\sigma_{eff}$,  we are allowed to use 
a representative selection of DPS processes with different final states and 
rather different kinematics. In particular,  we consider 
\textcolor{black}{the DPS production of double quarkonia and of high mass final
states, since this final state discrimination appears to be correlated with the 
extracted value of $\sigma_{eff}$. Therefore, we consider 
recent LHC analyses in which
$\sigma_{eff}$ is extracted} 
in the double $J/\Psi$ channel by the LHCb \cite{data8}, ATLAS \cite{data11} 
and by the authors of  Ref. \cite{data12} based on CMS data, in the 4-jets 
channel by ATLAS \cite{data6}, and in the  $W$+2 jets and same sign $WW$ 
channels analyzed by  CMS \cite{data9,data10}.
Results are presented in  Fig. \ref{distexp}, where 
the range of allowed mean partonic distance has been
calculated according  to Eq.  (\ref{main}) and displayed with  inner bars. 
%, i.e. \textcolor{blue}{$0.3<\sqrt{b^2}<0.95$ fm}.
\textcolor{black}{The theoretical uncertainty $\Delta$ associated with 
Eq. (\ref{main}), defined as the difference between the upper and lower limit of 
$\langle b^2\rangle$, 
parametrizes
the ignorance of the details of $f(k_\perp)$.}
The latter does depend linearly on $\sigma_{eff}$ so  
$\Delta$ gets smaller as $\sigma_{eff}$ decreases, a trend which can be 
observed in Fig. \ref{distexp}. Taking into account the experimental 
uncertainties associated with the $\sigma_{eff}$  extraction and adding them in 
quadrature to the theoretical ones, we obtain the outer error bars. We conclude 
that,  by using the extracted values of $\sigma_{eff}$ and their corresponding 
errors,  our estimate of the allowed range of $\langle b^2\rangle$ via Eq. 
(\ref{main}) is  dominated by the theoretical uncertainty, a conclusion that 
comes as no surprise since our result is obtained without assuming any detailed 
knowledge of the shape of $f(k_\perp)$.
It is worth noticing that the 
upper limit on the partonic distance for $\sigma_{eff}<20 \, $ mb is 
substantially smaller than the electromagnetic radius of the proton. This is 
{\it a posteriori} confirmation that measured values  
of $\sigma_{eff}$ are not compatible with trivial expectations based on 
geometrical considerations, and 
%\textcolor{blue}{and then the maximimum allowed geometrical value of $\langle b^2 \rangle$, i.e. 2$r_p$, being $r_p$ the proton radius. As one can notice, in Fig. \ref{distexp}, the y-axes range goes from $0$ to  2$r_p$ so that one can realize the reacness of information that can be obtained from  this analysis togheter with results of studies of $\sigma_{eff}$. Naturally, as one can see from Eq. (\ref{main}), the theoretical error on the evaluation of the main partonic distance is as small as the $\sigma_{eff}$ is small}. 
they directly point to dynamical correlation effects in the proton;
see the discussion in \cite{trele}.

In particular,
%in the kinematic region of recent $\sigma_{eff}$ experimental extractions, 
we have found a  minimum for the distance 
\textcolor{black}{
in the range $0.2 < \sqrt{\langle b^2\rangle_{min}} < 0.35$ {fm}, }
which is driven by $\sigma_{eff}$ 
extracted from processes involving heavy quarkonia pairs in the final state. 
On the other hand, 
%, due to experimental uncertenties, 
the maximum varies  in the range $0.6<\sqrt{\langle b^2\rangle_{max}}<0.95$ {fm} 
and  is driven by $\sigma_{eff}$  
extracted from processes involving electroweak bosons and/or jets.
 We point out that our mathematical approach 
works even if $\sigma_{eff}$ is not constant among different processes since it 
is sufficient that Eq. (\ref{sap}) holds.
{Therefore DPS measurements with final states whose production is dominated by 
distinct flavor species
will, hopefully,  allow the investigation of the flavor dependence 
of $\sigma_{eff}$ and consequently of the  effective form factor.}

%\vspace{1cm}
\section{Conclusions}
\label{conclusions}
In the present paper we have presented a method which allows us to convert 
the information encoded in $\sigma_{eff}$, a derived quantity often used in
experimental analyses to characterize the DPS cross section, into information 
on  
the partonic proton structure. The procedure used here makes a number of 
assumptions which are close to the ones frequently used in experimental 
analyses. Furthermore, we make no use of the detailed knowledge of the 
transverse form factor and only assume its reasonable
behavior at very large $k_\perp$ guided by studies on the standard proton form
factors.
In this way, the experimentally extracted $\sigma_{eff}$ can be directly cast
into a range of mean distances characterizing the interacting parton pair. 

%\vspace{1cm}
\section*{Acknowledgments}
This work was supported in part by Mineco under
Contracts No. FPA2013-47443-C2-1-P, Mineco and UE
Feder under Contracts No. FPA2016-77177-C2-1-P,
GVA-PROMETEOII/2014/066 and SEV-2014-0398.
We warmly thank Sergio Scopetta, Marco Traini, and
Vicente Vento for many useful discussions and comments
on the manuscript.


\begin{thebibliography}{50}
\bibitem{paver}
%\cite{Paver:1982yp}
%\bibitem{Paver:1982yp} 
  N.~Paver and D.~Treleani,
  %``Multi - Quark Scattering and Large $p_T$ Jet Production in Hadronic 
%Collisions,''
  Nuovo Cimento Soc. Ital. Fis. {\bf 70A}, 215 (1982).
%  doi:10.1007/BF02814035
  %%CITATION = doi:10.1007/BF02814035;%%
  %196 citations counted in INSPIRE as of 23 Oct 2017

%\cite{Sjostrand:1986ep}
\bibitem{Sjostrand:1986ep}
  T.~Sjostrand and M.~van Zijl,
  %``Multiple Parton-parton Interactions in an Impact Parameter Picture,''
  Phys.\ Lett.\ B {\bf 188}, 149 (1987); 
%\cite{Sjostrand:1987su}
%\bibitem{Sjostrand:1987su}
%  T.~Sjostrand and M.~van Zijl,
  %``A Multiple Interaction Model for the Event Structure in Hadron Collisions,''
  Phys.\ Rev.\ D {\bf 36}, 2019 (1987).
  %doi:10.1016/0370-2693(87)90722-2
  %%CITATION = doi:10.1016/0370-2693(87)90722-2;%%
  %78 citations counted in INSPIRE as of 05 Dec 2017


  
  %\cite{Diehl:2011yj}
\bibitem{ww_3_1} 
%\cite{Diehl:2011tt}
%\bibitem{Diehl:2011tt} 
  M.~Diehl and A.~Schafer,
  %``Theoretical considerations on multiparton interactions in QCD,''
  Phys.\ Lett.\ B {\bf 698}, 389 (2011); 
%  doi:10.1016/j.physletb.2011.03.024
%  [arXiv:1102.3081 [hep-ph]].
  %%CITATION = doi:10.1016/j.physletb.2011.03.024;%%
  %96 citations counted in INSPIRE as of 16 Feb 2017
M.~Diehl, D.~Ostermeier and A.~Schafer
%``Elements of a theory for multiparton interactions in QCD",
JHEP {\bf 03}, 089 (2012).
%[arXiv:1111.0910 [hep-ph]].
  %%CITATION = ARXIV:1111.0910;%%
  %19 citations counted in INSPIRE as of 26 Feb 2013

 \bibitem{Calucci:1999yz} 
  G.~Calucci and D.~Treleani, 
  %``Proton structure in transverse space and the effective cross-section,''
  Phys.\ Rev.\ D {\bf 60}, 054023 (1999)
%  doi:10.1103/PhysRevD.60.054023
%  [hep-ph/9902479].
  %%CITATION = doi:10.1103/PhysRevD.60.054023;%%
  %65 citations counted in INSPIRE as of 23 Oct 2017
  
  
%\bibitem{LH3}
  %\cite{Manohar:2012jr}
%\bibitem{Manohar:2012jr} 
 % A.~V.~Manohar and W.~J.~Waalewijn,
  %``A QCD Analysis of Double Parton Scattering: Color Correlations, 
%Interference Effects and Evolution,''
 % Phys.\ Rev.\ D {\bf 85}, 114009 (2012) \textcolor{blue}{TOGLIERE}
%  doi:10.1103/PhysRevD.85.114009
%  [arXiv:1202.3794 [hep-ph]].
  %%CITATION = doi:10.1103/PhysRevD.85.114009;%%
  %74 citations counted in INSPIRE as of 26 Oct 2017
  
  
%      \bibitem{plb_5}
  %\cite{Bansal:2014paa}
%\bibitem{bansal} 
%  S.~Bansal {\it et al.},
  %``Progress in Double Parton Scattering Studies,''
%arXiv:1410.6664 [hep-ph]~.
  %%CITATION = ARXIV:1410.6664;%%
  %5 citations counted in INSPIRE as of 24 Apr 2015
% \textcolor{red}{Siamo sicuri che vogliamo citare dei proceedings?}
    
%  \bibitem{LH5}
  %\cite{Szczurek:2015iva}
%\bibitem{Szczurek:2015iva} 
  %A.~Szczurek,
  %``Double parton scattering at high energies,''
  %Acta Phys.\ Polon.\ B {\bf 46}, no. 7, 1415 (2015) \textcolor{blue}{TOGLIERE}
%  doi:10.5506/APhysPolB.46.1415
%  [arXiv:1504.06491 [hep-ph]].
  %%CITATION = doi:10.5506/APhysPolB.46.1415;%%
  %9 citations counted in INSPIRE as of 26 Oct 2017
  
\bibitem{Mel_19}
%\cite{Chang:2012nw}
%\bibitem{Chang:2012nw} 
  H.~M.~Chang, A.~V.~Manohar and W.~J.~Waalewijn,
  %``Double Parton Correlations in the Bag Model,''
  Phys.\ Rev.\ D {\bf 87}, no. 3, 034009 (2013).
%  doi:10.1103/PhysRevD.87.034009
%  [arXiv:1211.3132 [hep-ph]].
  %%CITATION = doi:10.1103/PhysRevD.87.034009;%%
  %27 citations counted in INSPIRE as of 04 Aug 2016 
 
 \bibitem{noiold}
    %\cite{Rinaldi:2013vpa}
%\bibitem{Rinaldi:2013vpa} 
  M.~Rinaldi, S.~Scopetta and V.~Vento,
  %``Double parton correlations in constituent quark models,''
  Phys.\ Rev.\ D {\bf 87}, 114021 (2013)
%  doi:10.1103/PhysRevD.87.114021
%  [arXiv:1302.6462 [hep-ph]].
  %%CITATION = doi:10.1103/PhysRevD.87.114021;%%
  %35 citations counted in INSPIRE as of 10 Jan 2018
    
    
       \bibitem{noi1}%\cite{Rinaldi:2014ddl}
%\bibitem{Rinaldi:2014ddl} 
  M.~Rinaldi, S.~Scopetta, M.~Traini and V.~Vento,
  %``Double parton correlations and constituent quark models: a Light Front approach to the valence sector,''
  JHEP {\bf 12}, 028 (2014)
%  doi:10.1007/JHEP12(2014)028
%  [arXiv:1409.1500 [hep-ph]].
  %%CITATION = doi:10.1007/JHEP12(2014)028;%%
  %25 citations counted in INSPIRE as of 10 Jan 2018


    
  

     \bibitem{noij2}
  %\cite{Rinaldi:2016jvu}
%\bibitem{Rinaldi:2016jvu} 
  M.~Rinaldi, S.~Scopetta, M.~C.~Traini and V.~Vento,
  %``Correlations in Double Parton Distributions: Perturbative and 
%Non-Perturbative effects,''
  JHEP {\bf 10}, 063 (2016)
%  doi:10.1007/JHEP10(2016)063
%  [arXiv:1608.02521 [hep-ph]].
  %%CITATION = doi:10.1007/JHEP10(2016)063;%%
  %9 citations counted in INSPIRE as of 06 Oct 2017
  

  
  
%\bibitem{noi240}
%\cite{Diehl:2014vaa}
%\bibitem{Diehl:2014vaa} 
%  M.~Diehl, T.~Kasemets and S.~Keane,
  %``Correlations in double parton distributions: effects of evolution,''
%  JHEP {\bf 05}, 118 (2014)
 % doi:10.1007/JHEP05(2014)118
 % [arXiv:1401.1233 [hep-ph]].
  %%CITATION = doi:10.1007/JHEP05(2014)118;%%
  %32 citations counted in INSPIRE as of 23 Oct 2017

  

  
  
  
%   \bibitem{ww_23} 
%  A.~M.~Snigirev,
  %``Double parton distributions in the leading logarithm approximation of 
%perturbative QCD,''
%  Phys.\ Rev.\ D {\bf 68}, 114012 (2003).
  %doi:10.1103/PhysRevD.68.114012
  %[hep-ph/0304172].
  %%CITATION = doi:10.1103/PhysRevD.68.114012;%%
  %76 citations counted in INSPIRE as of 09 Feb 2017
% \textcolor{blue}{se afferma che l'effetto non omogeneo non è enorme si può tenere. ORa controllo anche io}



  
  
  


%\cite{Blok:2010ge}
%\bibitem{Blok:2010ge}
%  B.~Blok, Y.~Dokshitzer, L.~Frankfurt and M.~Strikman,
  %``The Four jet production at LHC and Tevatron in QCD,''
%  Phys.\ Rev.\ D {\bf 83} (2011) 071501
  
  
%  \bibitem{strikman}
%\bibitem{blok_2}
%\cite{Blok:2013bpa}
%  B.~Blok, Y.~Dokshitzer, L.~Frankfurt and M.~Strikman,
  %``Perturbative QCD correlations in multi-parton collisions,''
%  Eur.\ Phys.\ J.\ C {\bf 74}, 2926 (2014).
%  [arXiv:1306.3763 [hep-ph]].
  %%CITATION = ARXIV:1306.3763;%%
  %22 citations counted in INSPIRE as of 09 Feb 2015
  
  
  

  
  

  
  
%  \bibitem{Mel_11}
%H.~Jung \textsl{\& al.}, Proceedings, 7th International Workshop on Multiple
%                        Partonic Interactions at the LHC (MPI@LHC 2015),
%https://bib-pubdb1.desy.de/record/297386, DESY-PROC-2016-01
% \textcolor{red}{Siamo sicuri che vogliamo citare dei proceedings?}
  
  
%  \bibitem{uni1}
  %\cite{Gaunt:2010pi}
%\bibitem{Gaunt:2010pi} 
%  J.~R.~Gaunt, C.~H.~Kom, A.~Kulesza and W.~J.~Stirling,
  %``Same-sign W pair production as a probe of double parton scattering at the 
%LHC,''
%  Eur.\ Phys.\ J.\ C {\bf 69}, 53 (2010). \textcolor{blue}{TOGLIERE}
%  doi:10.1140/epjc/s10052-010-1362-y
%  [arXiv:1003.3953 [hep-ph]].
  %%CITATION = doi:10.1140/epjc/s10052-010-1362-y;%%
  %114 citations counted in INSPIRE as of 07 Nov 2017
  
%  \bibitem{uni2}
  %\cite{Echevarria:2015ufa}
%\bibitem{Echevarria:2015ufa} 
%  M.~G.~Echevarria, T.~Kasemets, P.~J.~Mulders and C.~Pisano,
  %``Polarization effects in double open-charm production at LHCb,''
%  JHEP {\bf 1504}, 034 (2015).
%  doi:10.1007/JHEP04(2015)034
%  [arXiv:1501.07291 [hep-ph]].
  %%CITATION = doi:10.1007/JHEP04(2015)034;%%
  %20 citations counted in INSPIRE as of 07 Nov 2017
%  \textcolor{blue}{Pubblicità, nel senso che anche Mulder usa questi approcci fattorizzati, giusto per dare più credibilità diciamo niente di più.}
%  \textcolor{red}{Ho riletto questo lavoro sulla polarizazzione in DPS nel canale ccbar: è pure peggio dei lavori dei cracoviani. Della serie faccio un lavoro, tanto non mi costa nulla. }
    
    
 % \bibitem{noiww}
  %\cite{Ceccopieri:2017oqe}
%\bibitem{Ceccopieri:2017oqe} 
 % F.~A.~Ceccopieri, M.~Rinaldi and S.~Scopetta,
  %``Parton correlations in same-sign $W$ pair production via double parton 
%scattering at the LHC,''
 % Phys.\ Rev.\ D {\bf 95}, no. 11, 114030 (2017) \textcolor{blue}{TOGLIERE}
 % doi:10.1103/PhysRevD.95.114030
 % [arXiv:1702.05363 [hep-ph]].
  %%CITATION = doi:10.1103/PhysRevD.95.114030;%%
  %4 citations counted in INSPIRE as of 23 Oct 2017
  
  
%    \bibitem{cao}
%\cite{Cao:2017bcb}
%\bibitem{Cao:2017bcb} 
%  Q.~H.~Cao, Y.~Liu, K.~P.~Xie and B.~Yan,
  %``Double Parton Scattering of Weak Gauge Boson Productions at the 13 TeV and 
%100 TeV Proton-Proton Colliders,''
%  arXiv:1710.06315 [hep-ph].
  %%CITATION = ARXIV:1710.06315;%%  
%  \textcolor{red}{TOGLIERE, questo se non  ricordo male è il lavoro cinese gemello del nostro sul WW }
  
  
  
  

  
  
  
  
    \bibitem{noir}
%\cite{Rinaldi:2016mlk}
%\bibitem{Rinaldi:2016mlk} 
  M.~Rinaldi and F.~A.~Ceccopieri,
  %``Relativistic effects in model calculations of double parton distribution 
%function,''
  Phys.\ Rev.\ D {\bf 95}, no. 3, 034040 (2017)
%  doi:10.1103/PhysRevD.95.034040
%  [arXiv:1611.04793 [hep-ph]].
  %%CITATION = doi:10.1103/PhysRevD.95.034040;%%
  %6 citations counted in INSPIRE as of 23 Aug 2017
  
    
  
  
  
  
  

    
    %  \bibitem{noiprd}
  %\cite{Rinaldi:2013vpa}
%\bibitem{Rinaldi:2013vpa} 
 % M.~Rinaldi, S.~Scopetta and V.~Vento,
  %``Double parton correlations in constituent quark models,''
 % Phys.\ Rev.\ D {\bf 87}, 114021 (2013)  \textcolor{blue}{TOGLIERE}
%  doi:10.1103/PhysRevD.87.114021
%  [arXiv:1302.6462 [hep-ph]].
  %%CITATION = doi:10.1103/PhysRevD.87.114021;%%
  %34 citations counted in INSPIRE as of 23 Oct 2017
  
  



  
    %\cite{Alitti:1991rd}
%\bibitem{data0} 
%  J.~Alitti {\it et al.}  [UA2 Collaboration],
  %``A Study of multi - jet events at the CERN anti-p p collider and a search 
%for double parton scattering,''
%  Phys.\ Lett.\ B {\bf 268}, 145 (1991).
  %%CITATION = PHLTA,B268,145;%%
  %128 citations counted in INSPIRE as of 17 juin 2015


 \bibitem{data8}
 %\cite{Aaij:2016bqq}
%\bibitem{Aaij:2016bqq}
  R.~Aaij {\it et al.} [LHCb Collaboration],
  %``Measurement of the J/$\psi$ pair production cross-section in pp collisions 
%at $ \sqrt{s}=13 $ TeV,''
  JHEP {\bf 06} (2017) 047
   Erratum: [JHEP {\bf 10} (2017) 068].



  \bibitem{data6}
  %\cite{Aaboud:2016dea}
%\bibitem{ATLAS_2016_4jet}
  M.~Aaboud {\it et al.} [ATLAS Collaboration],
  %``Study of hard double-parton scattering in four-jet events in pp collisions 
%at $ \sqrt{s}=7 $ TeV with the ATLAS experiment,''
  JHEP {\bf 11}, 110 (2016).




 
% \bibitem{data7}
 %\cite{CMS:2017jwx}
%\bibitem{CMS:2017jwx} 
%  CMS Collaboration [CMS Collaboration],
  %``Measurement of double parton scattering in same-sign WW production in p-p 
%collisions at $\sqrt{s}=13~\mathrm{TeV}$ with the CMS experiment,''
%  CMS-PAS-FSQ-16-009.
  %%CITATION = CMS-PAS-FSQ-16-009;%%
  %1 citations counted in INSPIRE as of 23 Oct 2017
% \textcolor{red}{Questa è la nota di CMS sul WW. Togliere?}
 
 


 
 \bibitem{data9}
 %\cite{Chatrchyan:2013xxa}
%\bibitem{Chatrchyan:2013xxa} 
  S.~Chatrchyan {\it et al.} [CMS Collaboration],
  %``Study of double parton scattering using W + 2-jet events in proton-proton 
%collisions at $\sqrt{s}$ = 7 TeV,''
  JHEP {\bf 03}, 032 (2014)
%  doi:10.1007/JHEP03(2014)032
%  [arXiv:1312.5729 [hep-ex]].
  %%CITATION = doi:10.1007/JHEP03(2014)032;%%
  %118 citations counted in INSPIRE as of 23 Nov 2017
 
 
%\cite{Sirunyan:2017hlu}
%\bibitem{Sirunyan:2017hlu}
 \bibitem{data10}
  A.~M.~Sirunyan {\it et al.} [CMS Collaboration],
  %``Constraints on the double-parton scattering cross section from same-sign W 
%boson pair production in proton-proton collisions at $ \sqrt{s}=8 $ TeV,''
  JHEP {\bf 1802}, 032 (2018)
%  doi:10.1007/JHEP02(2018)032
%  [arXiv:1712.02280 [hep-ex]].
  %%CITATION = doi:10.1007/JHEP02(2018)032;%%
  %1 citations counted in INSPIRE as of 30 Apr 2018
  
  
  
  \bibitem{data11}
  %\cite{Aaboud:2016fzt}
%\bibitem{Aaboud:2016fzt} 
  M.~Aaboud {\it et al.} [ATLAS Collaboration],
  %``Measurement of the prompt J/ $\psi $ pair production cross-section in pp collisions at $\sqrt{s} = 8$  TeV with the ATLAS detector,''
  Eur.\ Phys.\ J.\ C {\bf 77}, no. 2, 76 (2017)
%  doi:10.1140/epjc/s10052-017-4644-9
%  [arXiv:1612.02950 [hep-ex]].
  %%CITATION = doi:10.1140/epjc/s10052-017-4644-9;%%
  %19 citations counted in INSPIRE as of 17 Jan 2018
  
  
  \bibitem{data12}
 %\cite{Lansberg:2014swa}
%\bibitem{Lansberg:2014swa} 
  J.~P.~Lansberg and H.~S.~Shao,
  %``J/ψ -pair production at large momenta: Indications for double parton scatterings and large α$_s^5$ contributions,''
  Phys.\ Lett.\ B {\bf 751}, 479 (2015)
%  doi:10.1016/j.physletb.2015.10.083
%  [arXiv:1410.8822 [hep-ph]].
  %%CITATION = doi:10.1016/j.physletb.2015.10.083;%%
  %45 citations counted in INSPIRE as of 17 Jan 2018
 
 
   
   \bibitem{noiplb1}
  %\cite{Rinaldi:2015cya}
%\bibitem{Rinaldi:2015cya} 
  M.~Rinaldi, S.~Scopetta, M.~Traini and V.~Vento,
  %``Double parton scattering: a study of the effective cross section within a 
%Light-Front quark model,''
  Phys.\ Lett.\ B {\bf 752}, 40 (2016)
%  doi:10.1016/j.physletb.2015.11.031
%  [arXiv:1506.05742 [hep-ph]].
  %%CITATION = doi:10.1016/j.physletb.2015.11.031;%%
  %10 citations counted in INSPIRE as of 29 Aug 2017
  
 
 
    \bibitem{noi22}
  %\cite{Gaunt:2009re}
%\bibitem{Gaunt:2009re} 
  J.~R.~Gaunt and W.~J.~Stirling,
  %``Double Parton Distributions Incorporating Perturbative QCD Evolution and 
%Momentum and Quark Number Sum Rules,''
  JHEP {\bf 03}, 005 (2010).
%  doi:10.1007/JHEP03(2010)005
%  [arXiv:0910.4347 [hep-ph]].
  %%CITATION = doi:10.1007/JHEP03(2010)005;%%
  %139 citations counted in INSPIRE as of 23 Oct 2017 
 
 
   %\cite{Blok:2011bu}
\bibitem{ww_25_1}
  B.~Blok, , Y.~Dokshitzer, L.~Frankfurt and M.~Strikman,
  %``pQCD physics of multiparton interactions,''
 Phys.\ Rev.\ D {\bf 83} (2011) 071501;  Eur.\ Phys.\ J.\ C {\bf 72}, 1963 (2012); Eur.\ Phys.\ J.\ C {\bf 74}, 2926 (2014).
%  [arXiv:1106.5533 [hep-ph]].
  %%CITATION = ARXIV:1106.5533;%%
  %49 citations counted in INSPIRE as of 09 Feb 2015

 %\cite{Diehl:2015bca}
\bibitem{Diehl:2015bca}
  M.~Diehl, J.~R.~Gaunt, D.~Ostermeier, P.~Ploessl and A.~Schafer,
  %``Cancellation of Glauber gluon exchange in the double Drell-Yan process,''
  JHEP {\bf 01} (2016) 076
  %doi:10.1007/JHEP01(2016)076
  %[arXiv:1510.08696 [hep-ph]].
  %%CITATION = doi:10.1007/JHEP01(2016)076;%%
  %33 citations counted in INSPIRE as of 29 Nov 2017  
  
  
  \bibitem{add1}
  %\cite{Diehl:2017kgu}
%\bibitem{Diehl:2017kgu} 
  M.~Diehl, J.~R.~Gaunt and K.~Schonwald,
  %``Double hard scattering without double counting,''
  JHEP {\bf 1706}, 083 (2017)
%  doi:10.1007/JHEP06(2017)083
%  [arXiv:1702.06486 [hep-ph]].
  %%CITATION = doi:10.1007/JHEP06(2017)083;%%
  %8 citations counted in INSPIRE as of 21 Feb 2018

\bibitem{add2}
%\cite{Buffing:2017mqm}
%\bibitem{Buffing:2017mqm} 
  M.~G.~A.~Buffing, M.~Diehl and T.~Kasemets,
  %``Transverse momentum in double parton scattering: factorisation, evolution and matching,''
  JHEP {\bf 1801}, 044 (2018)
%  doi:10.1007/JHEP01(2018)044
%  [arXiv:1708.03528 [hep-ph]].
  %%CITATION = doi:10.1007/JHEP01(2018)044;%%
  %11 citations counted in INSPIRE as of 21 Feb 2018

  
  
  
  
  
%  \cite{Mekhfi:1983az}
\bibitem{Mekhfi:1983az}
  M.~Mekhfi,
  %``Multiparton Processes: An Application To Double Drell-yan,''
  Phys.\ Rev.\ D {\bf 32} (1985) 2371.

  
  

  
 %%%%%%%%%%%%%%%%%%%%%%%%%%%%%% 
  
 
  
  
  \bibitem{strikman3}
%\cite{Frankfurt:2002ka}
%\bibitem{Frankfurt:2002ka} 
  L.~Frankfurt and M.~Strikman,
  %``Two gluon form-factor of the nucleon and J / psi photoproduction,''
  Phys.\ Rev.\ D {\bf 66}, 031502 (2002)
 % doi:10.1103/PhysRevD.66.031502
 % [hep-ph/0205223].
  %%CITATION = doi:10.1103/PhysRevD.66.031502;%%
  %69 citations counted in INSPIRE as of 21 Dec 2017

  
  \bibitem{em1}
%\cite{Bernauer:2010wm}
%\bibitem{Bernauer:2010wm} 
  J.~C.~Bernauer {\it et al.} [A1 Collaboration],
  %``High-precision determination of the electric and magnetic form factors of the proton,''
  Phys.\ Rev.\ Lett.\  {\bf 105}, 242001 (2010)
%  doi:10.1103/PhysRevLett.105.242001
%  [arXiv:1007.5076 [nucl-ex]].
  %%CITATION = doi:10.1103/PhysRevLett.105.242001;%%
  %243 citations counted in INSPIRE as of 26 Dec 2017


\bibitem{em2}
%\cite{Andivahis:1994rq}
%\bibitem{Andivahis:1994rq} 
  L.~Andivahis {\it et al.},
  %``Measurements of the electric and magnetic form-factors of the proton from Q**2 = 1.75-GeV/c**2 to 8.83-GeV/c**2,''
  Phys.\ Rev.\ D {\bf 50}, 5491 (1994).
%  doi:10.1103/PhysRevD.50.5491
  %%CITATION = doi:10.1103/PhysRevD.50.5491;%%
  %362 citations counted in INSPIRE as of 26 Dec 2017


\bibitem{em3}
 %\cite{Gayou:2001qd}
%\bibitem{Gayou:2001qd} 
  O.~Gayou {\it et al.} [Jefferson Lab Hall A Collaboration],
  %``Measurement of G(Ep) / G(Mp) in polarized-e p ---> e polarized-p to Q**2 = 5.6-GeV**2,''
  Phys.\ Rev.\ Lett.\  {\bf 88}, 092301 (2002)
 % doi:10.1103/PhysRevLett.88.092301
 % [nucl-ex/0111010].
  %%CITATION = doi:10.1103/PhysRevLett.88.092301;%%
  %748 citations counted in INSPIRE as of 26 Dec 2017 
  
  \bibitem{brod}
 %\cite{Lepage:1980fj}
%\bibitem{Lepage:1980fj} 
  G.~P.~Lepage and S.~J.~Brodsky,
  %``Exclusive Processes in Perturbative Quantum Chromodynamics,''
  Phys.\ Rev.\ D {\bf 22}, 2157 (1980).
%  doi:10.1103/PhysRevD.22.2157
  %%CITATION = doi:10.1103/PhysRevD.22.2157;%%
  %3307 citations counted in INSPIRE as of 26 Dec 2017

  
  
%%%%%%%%%%%%%%%%%%
 


%\cite{Broniowski:2016trx}
%\bibitem{Mel_23}
%\bibitem{Broniowski:2016trx} 
%  W.~Broniowski, E.~Ruiz Arriola and K.~Golec-Biernat,
  %``Generalized Valon Model for Double Parton Distributions,''
%  Few Body Syst.\  {\bf 57}, no. 6, 405 (2016). \textcolor{blue}{TOGLIERE}
%  doi:10.1007/s00601-016-1087-z
%  [arXiv:1602.00254 [hep-ph]].
  %%CITATION = doi:10.1007/s00601-016-1087-z;%%
  %2 citations counted in INSPIRE as of 04 Aug 2016
  


  
  
 %\cite{Treleani:2007gi}
\bibitem{trele}
%\bibitem{Treleani:2007gi}
  D.~Treleani,
  %``Double parton scattering, diffraction and effective cross section,''
  Phys.\ Rev.\ D {\bf 76} (2007) 076006
  %doi:10.1103/PhysRevD.76.076006
  %[arXiv:0708.2603 [hep-ph]].
  %%CITATION = doi:10.1103/PhysRevD.76.076006;%%
  %46 citations counted in INSPIRE as of 07 Dec 2017 

  

  %doi:10.1103/PhysRevD.36.2019
  %%CITATION = doi:10.1103/PhysRevD.36.2019;%%
  %773 citations counted in INSPIRE as of 07 Dec 2017  
  

  
% \bibitem{datag1}
 %\cite{Binkley:1981kv}
%\bibitem{Binkley:1981kv} 
%  M.~E.~Binkley {\it et al.},
  %``J/psi Photoproduction from 60-GeV/c to 300-GeV/c,''
%  Phys.\ Rev.\ Lett.\  {\bf 48}, 73 (1982).
%  doi:10.1103/PhysRevLett.48.73
  %%CITATION = doi:10.1103/PhysRevLett.48.73;%%
  %196 citations counted in INSPIRE as of 21 Dec 2017
  
%  \bibitem{datag2}
  %\cite{Camerini:1975cy}
%\bibitem{Camerini:1975cy} 
%  U.~Camerini {\it et al.},
  %``Photoproduction of the psi Particles,''
%  Phys.\ Rev.\ Lett.\  {\bf 35}, 483 (1975).
%  doi:10.1103/PhysRevLett.35.483
  %%CITATION = doi:10.1103/PhysRevLett.35.483;%%
  %193 citations counted in INSPIRE as of 21 Dec 2017
    %doi:10.1103/PhysRevD.66.031502
 % [hep-ph/0205223].
  %%CITATION = doi:10.1103/PhysRevD.66.031502;%%
  %69 citations counted in INSPIRE as of 22 Dec 2017
  
%doi:10.1103/PhysRevD.83.071501
%  [arXiv:1009.2714 [hep-ph]].
  %%CITATION = doi:10.1103/PhysRevD.83.071501;%%
  %108 citations counted in INSPIRE as of 22 Dec 2017  
  
%%%%%%%%%%%%%%%%%
% FURTHER REFERENCES
%%%%%%%%%%%%%%%%%

%\cite{Goebel:1979mi}
%\bibitem{Goebel:1979mi}
%  C.~Goebel, F.~Halzen and D.~M.~Scott,
  %``Double {Drell-Yan} Annihilations in Hadron Collisions: Novel Tests of the Constituent Picture,''
%  Phys.\ Rev.\ D {\bf 22} (1980) 2789.



  
  
\end{thebibliography}
\end{document}